\documentclass{JHEP3} 

\usepackage{epsfig,multicol}

\title{String hamiltonian from generalized YM gauge theory in two dimensions}
\author{Florian Dubath, Simone Lelli, and Anna Rissone\\
D\'epartement de Physique Th\'eorique, Universit\'e de Gen\`eve,\\
24 quai Ernest-Ansermet, CH-1211 Gen\`eve 4\\
E-mail: \email{florian.dubath, simone.lelli, anna.rissone@physics.unige.ch}}

\preprint{}

\abstract{
Two dimensional $SU(N)$ Yang-Mills theory 
is known to be equivalent to a string
theory, as found by Gross in the large $N$ limit, using the $1/N$
expansion. 
Later it was found that even a generalized YM theory leads
to a string theory of the Gross type.
In the standard YM theory case, Douglas and others found the string
hamiltonian describing the propagation and the interactions of
states made of strings winding on a cylindrical space-time.
We address the problem of finding a similar hamiltonian for the
generalized YM theory. As in the standard case we start by writing
the theory as a theory of free fermions. Performing a bosonization, we
express the hamiltonian in terms of the modes of a 
bosonic field, that are interpreted as in the standard case as
creation and destruction operators for states of strings winding
around the cylindrical space-time.   
The result is similar to the standard hamiltonian, but with new kinds of
interaction vertices. 
}


\newcommand{\beq}{\begin{equation}}
\newcommand{\eeq}{\end{equation}}
\newcommand{\beqa}{\begin{eqnarray}}
\newcommand{\eeqa}{\end{eqnarray}}
\newcommand{\bit}{\begin{itemize}}
\newcommand{\eit}{\end{itemize}}
\newcommand{\refeq}[1] {(\ref{#1})}
\newcommand{\de}{\partial}

\newcommand{\la}{\lambda}
\newcommand{\al}{\alpha}

\newcommand{\bra}{\langle}
\newcommand{\ket}{\rangle}

\renewcommand\({\left(}
\renewcommand\){\right)}
\renewcommand\[{\left[}
\renewcommand\]{\right]}
\newcommand{\Bar}{\overline}
\newcommand{\Tilde}{\widetilde}
\renewcommand\leq{\leqslant}
\renewcommand\geq{\geqslant} 
\newcommand{\sss}{\scriptstyle}       
\newcommand{\tl}[2]{\begin{array}{c} #1 \\ #2 \end{array}}
\newcommand{\binom}[2]{\Big(\!\!\tl{#1}{#2}\!\!\Big)}

\newcommand\fverb{\setbox\pippobox=\hbox\bgroup\verb}
\newcommand\fverbdo{\egroup\medskip\noindent%
			\fbox{\unhbox\pippobox}\ }
\newcommand\fverbit{\egroup\item[\fbox{\unhbox\pippobox}]}
\newbox\pippobox

\begin{document}

\section{Introduction}

The idea of a relation between gauge theories and string theories is
very old, and a lot 
of progresses were made in this direction in the last years, with the
introduction of the Maldacena conjecture and the subsequent work.

A less recent approach to the problem is to study a
simpler model, the pure two dimensional $SU(N)$ Yang-Mills theory, which is
exactly solvable and can be described as a string theory
\cite{CMR}-\cite{minahan-ym-2}.   
In light of the recent developments, it is interesting to reconsider this
model, because it can be useful to gain a better
insight on the string/YM correspondence. 

Two-dimensional $SU(N)$ Yang-Mills theory 
does not contain gluons, nevertheless it is not trivial but there are
degrees of freedom related to the topology of the manifold \cite{
witten-ym}.
The interesting feature of this theory is that the partition function
can be exactly calculated on an arbitrary two dimensional manifold
\cite{migdal, rusakov}. 

The interpretation in terms of a string theory can be achieved in two
ways:
\bit
\item the partition function of the theory on a two dimensional
manifold $\Sigma_T$ for large $N$ can be 
developed in powers of $1/N$ and in this way it can be interpreted as
the partition function of a string theory with two-dimensional 
space-time $\Sigma_T$, coupling constant
$g_s = 1/N$ and string 
tension $\lambda = g^2 N$ (where $g$ is
the coupling constant of the gauge theory)\cite{gross1, gross2, gross3};  
\item the
two-dimensional Yang-Mills theory is equiva\-lent to a system of free
non-relativistic
fermions that in turn, for large $N$, can be rewritten as relativistic
ones; bosonizing one obtains a string description \cite{CMR, douglas,
douglas2}: one can thus calculate the string hamiltonian starting from
the Yang-Mills one.   
\eit
  
The starting point of our investigation is the fact that in two
dimensions the Yang-Mills action can be generalized to a class 
of theories; this is not possible in four dimensions where the
generalized theories are non-renormalizable. 
Since YM$_2$ is just a special point in this space of theories, it is
interesting to investigate if also the other theories
can be described in terms of strings. 

That this is indeed the case has been seen 
performing a $1/N$ expansion of the partition function analogous to that of
Gross and Taylor \cite{ganor, suga}. 
However the computation of the string hamiltonian by bosonizing
the fermionic description of the theory is a much more straightforward
way to obtain the string description; in this paper we
present the result we have obtained following this approach.

The paper is organized as follows:
In section \ref{generic casimirs} we introduce the generalized
theory, underlining the fact that the generalized hamiltonian in the
representation basis is simply a combination of higher Casimirs; then
we study its 
equivalence with a fermionic system, closely following the standard
case analysis and introducing the formulas that we have derived for the
calculation of the higher Casimirs. In the following sections we
focus on the quartic Casimir case, since we think that this example is
sufficient to understand the general features of the generalized
theories and to draw some conclusions.  
In particular in section \ref{quartic} we express the quartic Casimir
hamiltonian in terms
of non-relativistic fermionic fields, then in terms of
relativistic fermions and finally in terms of bosonic field modes,
thus obtaining the string hamiltonian. Then we discuss the result and
we make some considerations on the general case.  

\section{YM$_2$ generalized and description in terms of free fermions}
\label{generic casimirs}

To introduce the generalization of the Yang-Mills theory in two
dimensions, we need first to recall the definition of the standard theory.
The YM$_2$ action is
\beq\label{action1}
\mathcal{I}=-\frac{1}{2g^2}\int d^2x \sum_a F^{\mu \nu }_a F_{\mu \nu a}\,,
\eeq
where $a$ labels the generators of the group of internal symmetries.
In the following we consider first $G=U(N)$, then we extend the results
for the Yang-Mills theory for $SU(N)$.\footnote{We calculate the
corresponding string hamiltonian only for the $SU(N)$ theory, since in
this case (as for the standard theory \cite{ym2}) the result and the string
interpretation are simpler.}  

We consider the quantization on the cylinder with periodic spatial coordinate of period $L$.
As we work in two dimensions, $F^{\mu \nu }_a$ has a single degree of
freedom $F^{10}_a =E_a$; in the canonical way $E_a=-i g^2
\frac{\delta}{\delta A^1_a}$. Fixing the gauge by $A_0 =0$ the
equation of motion gives the constraint
$D_1 F_{10}\Psi=0$ that tells us 
that the wave-function $\Psi$ is simply a class function in the
holonomy variable $U=P \exp(i \int_0^L A_1 dx)$. 
The natural scalar product between wave-functions is
\beq
( \Psi_1 , \Psi_2 ) = \int_G d U \Psi_1^{*}(U) \Psi_2(U) \,,
\eeq
where $d U$ is the invariant (Haar) measure on gauge group $G$.

With this scalar product, a natural basis in the Hilbert space 
is the irreducible representations basis $\{|R\ket\}$,
with the corresponding wave-functions:
\beq
\label{irrbasis}
\bra U | R \ket = \chi_R (U) = \mbox{Tr}_R (U)\,.
\eeq

The hamiltonian takes the heat kernel form (see \cite{menotti}):
\beq\label{ham-heat-ker}
H=\int_0^Ldx\frac{1}{2g^2}\sum_a E_a E_a = - \int_0^L \frac{g^2}{2}\sum_a\frac{\delta}{\delta A^1_a}\frac{\delta}{\delta A^1_a}\,.
\eeq
Using $\frac{\delta}{\delta
A^1_a}\chi_R(U)=i \chi_R(T_a U)$ (where $\{T_a\}$ is  
a set of generators of the group) and the expression for the quadratic Casimir
$C_2 
=\sum_a T_a T_a$, in the basis \refeq{irrbasis} the
hamiltonian becomes: 
\beq\label{hamiltonian}
H = \frac{g^2}{2} L\, C_2(R)= \frac{\la L}{2N}C_2(R)\,,
\eeq
where $\la=N g^2$ (the t'Hooft coupling) is kept fixed in the $N\to \infty$
limit. 

The generalization of the theory is based on the fact that the partition
function of the standard theory  
\beq\label{part-func-old}
Z=\int\[DA^{\mu}\]e^{-\frac{1}{2g^2}\int d^2 x \sum_a
F^{\mu \nu }_a F_{\mu \nu a}} 
\eeq
is equivalent to the one built from the action
\beq\label{action2}
\mathcal{I}'=-\frac{1}{4}\int d^{2}x \( i\phi^a\epsilon^{\mu
\nu}F_{\mu \nu a}+\frac{g^2}{2}\sum_a \phi_a \phi_a\)\,, 
\eeq 
where $\phi$ is an auxiliary scalar field \cite{witten-ym}.
This formulation gives the possibility to generalize the YM$_2$
theory replacing in $\mathcal{I}'$ the term
$\frac{g^2}{2}\sum_a \phi_a \phi_a$ with 
terms of higher degrees in $\phi$ \cite{witten-ym}. Eliminating the auxiliary 
field from the partition function one can work out hamiltonians similar
to (\ref{ham-heat-ker}) (but with more than two derivatives). 
With such hamiltonians
one can build a generalized heat kernel equation \cite{ganor}. 
Writing this equation in terms of the holonomy variable $U$, 
one finally obtains the new hamiltonian looking like (\ref{hamiltonian})  
but with a linear combination of higher Casimirs instead of the
quadratic one:
\beq
H=\sum_k \frac{\la_k L}{N^{k-1}}C_k\,,
\eeq
where $\la_k = g^2_k N^{k-1}$ is the coupling to be held fixed in
the large $N$ limit (see \cite{ganor, suga}).

It is important to notice that this procedure gives
only symmetrized Casimirs (in the sense specified in
Sect.~\ref{calc-higher}), 
see \cite{ganor}. 

\subsection{Free fermions for a generic Casimir invariant}

To study the equivalence of the generalized YM theory with a system of
non-relativistic fermions, we first briefly summarize the method used in
the standard case \cite{parisi, douglas}.   

Since class functions are determined by their values on the maximal
torus ($U = \textrm{Diag} (e^{i \theta_1}, \ldots, e^{i \theta_N})$) and
they are invariant for exchange of the eigenvalues, they are totally
symmetric functions $\Psi(\theta_1,\ldots, \theta_N)$. The scalar
product becomes:
\beq
(\Psi_1, \Psi_2) = \frac{1}{N! (2 \pi)^N} \int \prod d \theta_i
\Tilde{\Delta}(\vec{\theta})^2 \Psi_1^*(\vec{\theta}) \Psi_2(\vec{\theta})\,,
\eeq  
where $\Tilde{\Delta}(\vec{\theta}) = \prod_{i < j} \sin \frac{\theta_i -
\theta_j}{2}$.
In the standard case  the hamiltonian in terms of the $\vec \theta$ is
\beq\label{ham-ferm}
H = \frac{\la}{2N} L \hat C_2 =\frac{\la}{2N} L
\frac{1}{\Tilde{\Delta}}\[\sum_i \(- 
\frac{\de^2}{\de \theta_i^2}\) - F_2\]
\Tilde{\Delta}\,,
\eeq
with $F_2 = \frac{N(N^2-1)}{12}$.

Looking at the hamiltonian and the scalar product, one can see that
it is better to use as wave-functions the anti-symmetric 
$\psi(\vec{\theta}) =  
\Tilde{\Delta}(\vec{\theta}) \Psi(\vec{\theta})$, obtaining in this way a
non-relativistic theory of $N$ free fermions on the circle.

The generic base state $|R\ket$ with associated Young diagram with row lengths
$h_i$ corresponds in this fermionic theory to the base state
$\psi_{\vec{n}}(\vec{\theta}) = \mbox{det}_{1 \leq i, j \leq N} \:
e^{i n_j \theta_i}$ with $n_j = h_j - j + n_F + 1$
(where $n_F = \frac{N - 1}{2}$ is the Fermi level).

From the eigenvalues equation for $\hat C_2$:
\beq
\hat C_2\Psi_R=C_2(R) \Psi_R\,, 
\eeq
we obtain the equation for $\psi$:
\beq
C_2(R)\psi_R = C_2(R) \Tilde \Delta \Psi_R= \Tilde \Delta \hat
C_2\Psi_R = -\de_i^2 (\Tilde \Delta \Psi_R)-F_2 \Tilde \Delta \Psi_R=
-\de_i^2 \psi_R-F_2 \psi_R
\eeq
that is
\beq
\label{C2 psiR}
\hat C_2'\psi_R\equiv -\de_i^2 \psi_R-F_2 \psi_R=\bigg[\sum_i
n_i^2-F_2\bigg]\psi_R=C_2(R)\psi_R\,. 
\eeq

Similarly for the higher Casimirs the same wave-functions $\Psi$
satisfy 
\beq
\hat C_k \Psi_R=C_k(R)\Psi_R\,,
\eeq
that in terms of $\psi=\Tilde \Delta \Psi$ becomes
\beq
\hat C_k'\psi_R=\Tilde \Delta \hat C_k 
\(\frac{1}{\Tilde \Delta}\psi_R\)=C_k(R)\psi_R\,,
\eeq 
i.e. an equation analogous to \refeq{C2 psiR}. 

We are not interested in writing the $\hat C_k$ operator, because we
can start directly from the eigenvalues of the fermionic hamiltonian 
$C_k(R)$, 
which are simply the Casimir operators in terms of the $n_i$'s.
In particular we focus on the theory with a quartic
Casimir, since it is the simplest generalization, sufficient to show
the new features of the general case. A cubic Casimir or a Casimir of
order $k=2m+1$ (odd) is not suitable since it has a leading
$n_i^k=n_i^{(2m+1)}$ term for $n_i\to -\infty$, thus the
eigenvalues of the hamiltonian are not bounded from below, and
therefore the theory is not quantizable.         

\subsection{Calculation of the higher Casimirs}
\label{calc-higher}
In this subsection we show how to write
$C_k(R)$ as a function of the $\{n_i\}$, and this is the starting
point for
the subsequent analysis as stated above.
To obtain this expression for  $C_k(R)$, we start from the higher
Casimir operators illustrated in \cite{CMR}, Sect. 4.10.
   
The $k$-th symmetrized Casimir operator for a representation of rank
$n$ of $U(N)$ (that is built 
from the direct product of $n$ copies of the fundamental
representation) is 
\beq
\widehat C_k=\sum_{i_1\dots i_{k}} \sum_{a_1 \dots a_k}
E^{(a_1}_{i_1 i_2} E^{a_2}_{i_2 i_3} \cdots E^{a_k)}_{i_{k} i_{1}}\,,  
\eeq 
where the $i$'s are in ${1,2..N}$, the $a$'s label the $n$ spaces and
$E^a_{ij}$ is the $(i,j)$-generator in the Lie algebra of the $a$-th 
space. $(..)$ means total symmetrization.\footnote{Non-symmetrized
Casimirs on one hand are excluded from YM$_2$ considerations as stated above, 
and on the other hand they give a string hamiltonian asymmetric under the
exchange $\al_n\leftrightarrow\Tilde\al_n$, thus a hamiltonian that
unnaturally distinguishes between a winding direction and the
opposite. This is again an hint of correspondence between the
Yang-Mills and the string theories.} 

The generators $E^a_{ij}$ are given by
\beq
(E^a_{ij})_{\alpha\beta}=\delta_{i\alpha}\delta_{j\beta}\,,
\eeq
and obey the following commutation rules
\beq
[E^a_{ij},E^b_{kl}]=0 \quad \textrm{if } a\neq b \qquad
[E^a_{ij},E^a_{kl}]=\delta_{jk} E^a_{il}-\delta_{il}E^a_{kj}\,.
\eeq   

We need the eigenvalue of the Casimir operator for a particular
representation $R$ of rank $n$. This can be found as $C_k(R)=\chi_R(\widehat
C_k)/d_R$. 

The first Casimir is
\beq
C_1=\sum_i \sum_a E^a_{ii}=n\,. 
\eeq
To write higher Casimirs, we use (see \cite{CMR}): 
\beq
\sum_{i_1\dots i_{k}} \sum_{
\begin{array}{c} 
\sss a_1 \dots a_k \\
\sss \forall i\neq j,\; a_i\neq a_j
\end{array}}
E^{a_1}_{i_1 i_2} E^{a_2}_{i_2 i_3} \cdots E^{a_k}_{i_{k} i_{1}}=P_k\,,  
\eeq  
where $P_k=k T_k$, and $T_k$ is the sum of the elements in the 
class of permutations in $S_n$ with a cycle
of length $k$ and the remainings of length 1.
 
When we evaluate this on a representation we obtain 
\beq
\gamma_k\equiv \frac{\chi_R(P_k)}{d_R}=k\frac{\chi_R(T_k)}{d_R}\,.
\eeq

Expanding the Casimir operators in terms of the $P_k$, it turns out
that they can be expressed as functions of $C_1$ and
some $\gamma_k$'s. 
For example for the quadratic Casimir (where the symmetrization is not
needed because we are tracing, so the operator is automatically
symmetrized) we reobtain the formula of the previous section: 
\beqa
{\hat C}_2&=&\sum_{i_1\, i_2} \sum_{a_1\, a_2}E^{a_1}_{i_1 i_2} E^{a_2}_{i_2
i_1}=\nonumber\\
&=&\sum_{i_1\, i_2} \sum_{a_1\neq a_2}E^{a_1}_{i_1 i_2} E^{a_2}_{i_2 i_1}+
\sum_{i_1\, i_2} \sum_{a}E^{a}_{i_1 i_2} E^{a}_{i_2 i_1} \quad
\Rightarrow \nonumber\\
C_2&=&\gamma_2+\sum_{i_1\, i_2} \sum_{a}\delta_{i_2 i_2} E^{a}_{i_1 i_1}=
\gamma_2+N C_1\,.
\eeqa
Isolating the various cases of equal and unequal indexes yields
the desired form for the Casimirs, expressed 
with $\gamma_2$,$\gamma_3$ etc..
For example we have for the 3rd non-symmetric Casimir:
\beq
C_3^{(non-sym)}=\gamma_3+2N \gamma_2+C_1(C_1-1)+N^2 C_1\,.
\eeq   
And the 3rd symmetric Casimir:
\beq
C_3=\gamma_3+\frac{3N}{2}\gamma_2+\frac{3}{2}C_1(C_1-1)+\frac{N^2+1}{2}C_1=
\gamma_3+\frac{3N}{2}\gamma_2+\frac{3}{2}C_1^2+\frac{N^2-2}{2}C_1\,.
\eeq
A very long calculation gives the 4th symmetric Casimir:
\beq
C_4=\gamma_4+2 N \gamma_3+\( 4 C_1+\frac{7
N^2}{6}-5\)\gamma_2+\frac{17 N}{6}C_1^2+\frac{N(N^2-12)}{6}C_1\,. 
\eeq

At this point we need the expression of these invariants in terms of the data
that specify a single representation, i.e. the free fermions momenta
$\{n_i\}$ that are related to the row lengths $\{h_i\}$ of the Young
diagram by $\{n_i=h_i+\frac{N+1}{2}-i\}$. Same intermediate
calculations are made with the set
$\{l_i=h_i+N-i=n_i+\frac{N-1}{2}\}$.

The $\{\gamma_k\}$ are expressed in terms of the $\{l_i\}$ (see the
App. \ref{gamma in l}, where a general method to calculate a
$\gamma_k$ with a given $k$ is explained):
\beqa
\label{gammalist}
\gamma_1&=&\sum_i l_i-\binom{N}{2}\nonumber\\
\gamma_2&=&\sum_i l_i^2-(2N-1)\sum_i l_i+\frac{N(N-1)(2N-1)}{3}\nonumber\\
\gamma_3&=&\sum_i l_i^3-3 N \sum_i l_i^2-3 \sum_{i<j}l_i l_j+
\frac{9N^2-9N+4}{2} \sum_i l_i+\\
&-&\frac{3}{8}N(N-1)(3N^2-3N+2)\nonumber\\
\gamma_4&=&\sum_i l_i^4-4 \sum_{i<j} (l_i^2 l_j+l_i
l_j^2)-(4N+2)\sum_i l_i^3+ (16N-8) \sum_{i<j}l_i l_j+ \nonumber\\
&+&(8N^2+3)\sum_i
l_i^2-\frac{2}{3}(2N-1)(8N^2-8N+9) \sum_i l_i+\nonumber\\
&+&\frac{4}{7}N(N-1)(2N-1)(4N^2-4N+7)\nonumber\,.
\eeqa 

With this expressions we arrive to the expressions for the Casimir
invariants in terms of $\{n_i\}$.
The quadratic Casimir is   
\beq
C_2^{U(N)}=N C_1+\gamma_2=\sum_i l_i^2-(N-1)\sum_i
l_i+\binom{N}{3}=\sum_i n_i^2-F_2\,, 
\eeq
while the quartic is
\beqa
C_4^{U(N)}&=&\sum_i l_i^4-2(N+1)\sum_i l_i^3-4\sum_{i<j} (l_i^2 l_j+l_i
l_j^2)+\nonumber\\
&+&\bigg(\frac{7N^2}{6}
+\frac{29 N}{6}-2+4\sum_j l_j\bigg)\sum_i l_i^2-4(2N-1)\Big(\sum_i l_i\Big)^2+
\\
&+&\bigg(\frac{47 N}{3}-8\bigg)\sum_{i<j}
l_i l_j-\binom{N}{2}\frac{17 N^3+17 N^2-108N+72}{180}\nonumber\\
&=&\sum_i n_i^4+\frac{3-2N^2}{6}\sum_i n_i^2-\frac{N}{6}(\sum_i
n_i)^2+\frac{N(N^2-1)(11N^2-9)}{720} \nonumber\,.
\eeqa 

Casimirs of $SU(N)$ come from the shift (notice that this assures
translation invariance of the $SU(N)$ Casimirs)\footnote{For a more
precise derivation see \cite{CMR}, Sect. 4.10}:
\beq
n_i\to n_i-\frac{\sum_j n_j}{N}\,,
\eeq 
From this substitution we have the 4th symmetric Casimir of $SU(N)$:

\beqa
\label{quartic Casimir in h}
C_4^{SU(N)}&=&\sum_i n_i^4-\frac{4 (\sum_i n_i)(\sum_j n_j^3)}{N}+\frac{2
N^2-3}{6N}(\sum_i n_i)^2+\\
&+&\bigg(\frac{6 (\sum_j
n_j)^2}{N^2}-\frac{2N^2-3}{6}\bigg) \sum_i n_i^2-\frac{3 (\sum_i
n_i)^3}{N^3}+\frac{N(N^2-1)(11 N^2-9)}{720}\nonumber\,.
\eeqa

More generally this procedure could produce any desired symmetric or
non-symmetric higher Casimir invariant for $U(N)$ (or $SU(N)$). We
will see that the quartic Casimir is sufficient to show many
peculiarities of the higher Casimir case.   

\section{String hamiltonian from the generalized YM$_2$ with Quartic Casimir}
\label{quartic}

In this section we repeat the steps that lead to the string
hamiltonian in the standard case, starting from the generalized
Yang-Mills theory with a quartic Casimir. The notations and the methods
are those of \cite{ym2}. 

As in the case of quadratic
Casimir, we write the YM theory as a system of free
non-relativistic fermions, then in the large N limit the fermionic theory
is written as a relativistic theory of fermions,
a $bc$ theory that can be bosonized. As in the standard case the bosonic
creation and destruction operators are interpreted as the operators that create
and destroy states of strings with a given winding on the cylinder.
 
First of all we introduce the second quantization
formalism for the non-relativistic fermions, by means of the operators
$B_n$ and $B^{\dagger}_n$ that 
destroy and create a fermion in the $n$-th level, 
with the anticommutation relations:
\beq\label{anticommutation}
\left\{B_n, B_m^{\dagger}\right\} = \delta_{n, m}\,,
\eeq
and the constraint that the number of particle is $N$:
\beq\label{constraint}
\sum_n B_n^{\dagger} B_n = N\,.
\eeq

The fundamental state of the theory must satisfy the relations:
\beqa
B_n | 0 \ket_F = 0 & \mbox{if} & |n| > n_F\label{vuotoF}\\
B_n^{\dagger} | 0 \ket_F = 0 & \mbox{if} & |n| \leq n_F\nonumber\,,
\eeqa
and the normal ordering $::$ is defined with respect to this vacuum
state.
 
With this normal ordering, the constraint (\ref{constraint}) becomes:
\beq\label{constraint2}
\sum_n :B_n^{\dagger} B_n: = 0\,.
\eeq
 
Using the prescription
\beq\label{f-to-s}
\sum_i n_i^k\to \sum_n n^k B_n^\dag B_n=\sum_n n^k :B_n^\dag B_n:+F_k\,,  
\eeq
with\footnote{We
have $F_{2l+1}\equiv 0$,
$F_2$ as above is $\frac{N(N^2-1)}{12}$ while
$F_4=\frac{N(N^2-1)(3N^2-7)}{240}$.} $F_k=\sum_{n=-n_F}^{+n_F} n^k$,  
we can write 
the quartic Casimir \refeq{quartic Casimir in h} as 
(we see that the pure constant terms cancel, as they should since
$H|0\ket_F=0$): 
\beqa
C_4&=&\sum_n n^4 :B_n^\dag B_n: - \frac{4}{N} \sum_n n :B_n^\dag
B_n: \sum_{n'} {n'}^3 :B_{n'}^\dag B_{n'}: +\nonumber 
\\
&+&\frac{6}{N} \Big(\sum_n n :B_n^\dag
B_n:\Big)^2 \sum_{n'} {n'}^2 :B_{n'}^\dag B_{n'}: +  
\frac{N^2 - 1}{2} \Big(\sum_n n :B_n^\dag B_n:\Big)^2 +\nonumber 
\\
&-& \frac{3}{N^3}
\Big(\sum_n n :B_n^\dag B_n:\Big)^4 + N \sum_n n^3 :B_n^\dag
B_n:+ \\  
&-&3 \Big(\sum_n n :B_n^\dag B_n:\Big) \Big(\sum_{n'} {n'}^2 :B_{n'}^\dag B_{n'}: \Big)+\nonumber
\\
&-&\frac{N(N^2 - 1)}{4} \sum_n n :B_n^\dag B_n: + \frac{2}{N} \Big(\sum_n n
:B_n^\dag B_n:\Big)^3+\nonumber
\\ &+& \frac{1}{2}\sum_n n^2 :B_n^\dag B_n: 
- \frac{1}{2N} \Big(\sum_n n :B_n^\dag B_n:\Big)^2\nonumber\,.
\eeqa 

\subsection{Relativistic fermions, $bc$ theory} 

As in \cite{ym2}, 
at large $N$, we can express the fermionic operators $B_n$ and
$B^\dag_n$ in terms of the $bc$ and $\Tilde b \Tilde
c$ fields as 
\beqa
b_n = B_{n_F + 1 + n},&\qquad&
c_n = B_{n_F + 1 - n}^{\dagger},\label{Bbc}\\
\Tilde{b}_n = B_{- n_F - 1 - n},&\qquad&
\Tilde{c}_n = B_{- n_F - 1 + n}^{\dagger}\nonumber\,.   
\eeqa
In the large $N$ limit the cut-off on the
modes of $b$, $c$, $\Tilde b$ and $\Tilde c$, can be removed, as the
missing contributions are 
non-perturbative (see \cite{ym2} for a discussion of this point); hence
here $n\in\{-\infty,+\infty\}$\footnote{Thus 
as in \cite{ym2} we can use the usual bosonization formulas. Here we
do not set to zero the modes $b_n, c_n$ with $|n|>n_F$, only for
simplicity in the notation.}.

The $bc$, $\Tilde b \Tilde c$ fields satisfy the anticommutation relations:
\beqa
\{c_n, b_m\} = & \{\Tilde{c}_n, \Tilde{b}_m\} & =
\delta_{n + m, 0}\label{anticomm-bc}\\
\{c_n, \Tilde{b}_m\} = & \{\Tilde{c}_n, b_m\} &  = 0\nonumber
\eeqa  
and the constraint:
\beq\label{vinc-bc}
\sum_{n} (:c_{- n} b_n: + :\Tilde{c}_{- n} \Tilde{b}_n:) = 0\,.
\eeq

From these operators we can build the fields:
\beqa
& & b(z)=\sum_{n} \frac{b_n}{z^{n+\la}},\qquad
c(z)=\sum_{n} \frac{c_n}{z^{n+1-\la}},\\
& & \Tilde b(\Bar z) = \sum_{n} \frac{\Tilde b_n}{\Bar z^{n+\la}}, \qquad
\Tilde c(\Bar z) = \sum_{n} \frac{\Tilde c_n}{\Bar z^{n+1-\la}}\,,
\eeqa
where $z$ is a complex variable, while $\lambda$ is a real parameter: we
consider it generic since it is a good check to verify that the  
hamiltonian does not depend on it (as it should, since it is an
auxiliary parameter).
From the anticommutation relations we can see that this is a $b
c$ and $\Tilde{b} \Tilde{c}$ theory (see \cite{polch}, sect. 2.7); 
moreover the vacuum $|0\ket_F$ is the ground state $|\downarrow\ket$.

We have to write expressions like \refeq{f-to-s} in terms of the $b,c$
modes:     
\beq
\sum_n n^k :B_n^\dag B_n :=\sum_{n} (n_F+1+n)^k \[:c_{-n}
b_{n}:+(-1)^k:\Tilde c_{-n} \Tilde b_{n}:\]\,.
\eeq
Using 
\beq
(n_F+1+n)^k=\sum_{i=0}^k n^i \(\frac{N+1}{2}\)^{k-i}\binom{\;k\;}{\;i\;}\,,
\eeq
we have to calculate sums like
\beq
\label{sumntoi}
\sum_n n^i :c_{-n} b_{n}: 
\eeq
(and similar for tilded fields). 
The \refeq{sumntoi} can be expressed easily 
as a linear combination of integrals of type:
\beq
W_{ab}\equiv \oint \frac{dz}{2\pi i} z^{a+b}:\de^a c(z) \de^b b(z):\,,
\eeq
since their expression in terms of the modes is 
\beq
\label{WabN}
W_{ab}=(-1)^b \sum_n \prod_{i=-a}^{b-1} (n+\la+i) :c_{-n} b_{n}:\,.
\eeq

Inverting \refeq{WabN} yields
\beqa
\sum_n n :c_{-n}b_{n}: &=& -W_{01}-\la W_{00} \nonumber\\
\sum_n n^2 :c_{-n}b_{n}: &=& W_{11}+(2\la-1) W_{01}+\la^2
W_{00}\nonumber\\
\sum_n n^3 :c_{-n}b_{n}: &=& W_{12}+3\la
W_{11}-(3\la^2-3\la+1)W_{01} -\la^3 W_{00}\\
\sum_n n^4 :c_{-n}b_{n}: &=&
W_{22}-2(2\la-1)W_{12}-(6\la^2+1)W_{11}+\nonumber\\ 
&+&(4\la^3-6\la^2+4\la-1)W_{01}+\la^4 W_{00}\nonumber\,.
\eeqa
The antichiral expression are the same, all with tilded objects.

At this point we have the expressions of $\sum_n n^k :B_n^\dag B_n:$
for $n=\{1,2,3,4\}$, as functions of $W_{00}$, $W_{01}$, $W_{11}$,
$W_{12}$ and $W_{22}$.

\subsection{Bosonization}

The last step is to bosonize the $b c$ and $\Tilde{b} \Tilde{c}$
theory in the standard way (see \cite{polch}, sect. 10.7). 
First we must notice
that to bosonize we need the conformal ordering instead of the
annihilation-creation one; in the $b c$ theory there is the relation:  
\beq
\label{conffromnorm}
:b(z)c(z'):_c=:b(z)c(z'):+\frac{z^{1-\la} z'^{\la-1}-1}{z-z'}
\eeq
(and similar for $\Tilde b \Tilde c$).

We can now bosonize in terms of the field $X(z, \overline{z}) = X_L(z) 
+ X_R(\overline{z})$:
\beqa
b(z) & = & :e^{i X_L(z)}:_c\nonumber\\
c(z) & = & :e^{- i X_L(z)}:_c\label{rel-bos}\\
\Tilde{b}(\overline{z}) & = & :e^{i X_R(\overline{z})}:_c\nonumber\\
\Tilde{c}(\overline{z}) & = & :e^{- i X_R(\overline{z})}:_c\nonumber\,.
\eeqa
For $\lambda = 1/2$ we obtain a scalar free field theory; for
$\lambda$ generic this is a linear dilaton CFT.
 
We can expand the bosonic field in modes:
\beqa
\partial X_L(z) & = & i \sum_{n \in \mathbb{Z}}
\frac{\alpha_n}{z^{n + 1}}\label{svil-bos}\\
\overline{\partial} X_R(\overline{z}) & = & i \sum_{n \in \mathbb{Z}}
\frac{\Tilde{\alpha}_n}{\overline{z}^{n + 1}}\nonumber\,,
\eeqa
which satisfy the commutation relations:
\beqa
\[\alpha_m, \alpha_n\] & = & \[\Tilde{\alpha}_m, \Tilde{\alpha}_n\] = m
\delta_{m + n, 0}\\
\[\alpha_m, \Tilde{\alpha}_n\] & = & 0\nonumber\,.
\eeqa

Following the Gross-Taylor interpretation of the standard Yang-Mills partition
function in 
terms of a string theory \cite{gross1, gross2, gross3}
and the subsequent construction of the hamiltonian by Minahan and Polychronakos
\cite{minahan-ym}, 
the modes $\alpha_n$ with $n > 0$ ($n < 0$) can be
interpreted as operators of annihilation (creation) of a string
winding $n$ times around the spatial circle in the same sense as the
orientation; similarly the modes $\Tilde{\alpha}_n$ describe strings
winding in the opposite sense.
Notice that this interpretation does not depend on $\lambda$ as it
should be; we will
check that the same happens for our hamiltonian from the quartic Casimir. 

Finally the constraint (\ref{vinc-bc}) in terms of the bosonic modes
is:
\beq\label{vinc-bos}
\alpha_0 + \Tilde{\alpha}_0 = 2 (\lambda - 1)\,.
\eeq

\vspace{0.8cm}\mbox{}\\
To bosonize, 
we need first to express $W_{ab}$ in terms of
\beq
W_{ab}^c\equiv \oint \frac{dz}{2\pi i} z^{a+b}:\de^a c(z) \de^b b(z):_c\,,
\eeq
where the ordering is the conformal ordering of the $bc$ CFT.
The relation between the two types of $W$ can be derived from
\refeq{conffromnorm} and is 
\beqa
W_{ab}&=&\oint \frac{dz}{2\pi i} z^{a+b}:\de^a c(z) \de^b
b(z):=W_{ab}^c+\oint \frac{dz}{2\pi i} z^{a+b} \lim_{z'\to z}\de^a
\de'^b\frac{z'^{1-\la} z^{\la-1}-1}{z'-z}=\nonumber \\
&=&W_{ab}^c+S_{ab}\,. 
\eeqa
In particular the shifts used in the following are
\beqa
& & S_{00}=-(\la-1), \qquad S_{01}=\frac{\la(\la-1)}{2},
\qquad S_{11}=\frac{\la(\la-1)(\la-2)}{3} \nonumber \\
& & S_{12}=-\frac{\la(\la^2-1)(\la-2)}{4}, \qquad 
S_{22}=-\frac{\la(\la^2-1)(\la-2)(\la-3)}{5}\,.
\eeqa

The main step is to write down $W_{ab}^c$, and then
$W_{ab}$, in terms of the modes of
$X_L(z)$ and $X_R(\Bar z)$ 
as in the quadratic case. 

Bosonization is carried out using the following relations between the
fermionic and bosonic fields
\beqa
:c(z)\de b(z):_c&=&\frac{1}{2}:(\de
X_L(z))^2:_c-\frac{i}{2}:\de^2 X_L(z):_c\nonumber\\
:\de c(z) \de b(z):_c&=&-\frac{i}{3}:(\de X_L)^3:_c-\frac{i}{6}:\de^3
X_L:_c\nonumber\\
:\de c(z)\de^2 b(z):_c&=&\frac{1}{4}:(\de
X_L(z))^4:_c-\frac{i}{2}:(\de X_L(z))^2
\de^2X_L(z):_c+\nonumber\\ &+&\frac{1}{4}:(\de^2
X_L(z))^2:_c-\frac{i}{12}:\de^4 X_L(z):_c\\
:\de^2 c(z) \de^2 b(z):_c&=&-\frac{i}{5}:(\de
X_L)^5:_c-i:\de X_L(z)(\de^2
X_L(z))^2:_c-\frac{i}{30}:\de^5 X_L(z):_c\nonumber
\eeqa
(and the similar relations between $\Tilde b, \Tilde c$ and $X_R$).
From these and 
\beqa
L_0=\sum n:c_{-n}b_n:_c\,,
\eeqa
the normal ordered integrals are easily calculated in terms of
the bosonic modes, giving
\beqa
W_{00}&=&\al_0-(\la-1)\nonumber\\
W_{01}&=&-L_0+\frac{\la(\la-1)}{2}-\la \al_0\nonumber\\ 
W_{11}&=&\frac{\al_0(1-\al_0^2)}{3}+\frac{\la(\la-1)(\la-2)}{3}-2\al_0 {\cal N}-A_3\\
W_{12}&=&\frac{\al_0(\al_0^2-1)(\al_0+2)}{4}-\frac{\la(\la^2-1)(\la-2)}{4}+
\Big(3\al_0(\al_0+1)-\frac{1}{2}\Big){\cal N} +\nonumber\\
&+&\frac{1}{2}A_{2n^2}+(3\al_0+\frac{3}{2})A_3 +A_4 \nonumber\\
W_{22}&=&\frac{\al_0(\al_0^2-1)(\al_o^2-4)}{5}
-\frac{\la(\la^2-1)(\la-2)(\la-3)}{5} +\nonumber \\
&+&2\al_0((2\al_0^2-3){\cal
N}+A_{2n^2}) -3\Big(A_3-\frac{A_{3n^2}}{2}\Big)+6\al_0^2 A_3+4 \al_0 A_4+A_5\nonumber\,,
\eeqa
with
\beqa
{\cal N}&=&A_2=\sum_{n> 0} \al_{-n}\al_{+n}\nonumber\\
A_{2n^2}&=&\sum_{n> 0} n^2\al_{-n}\al_{+n}\nonumber\\
A_3&=&\sum_{\tl{\sss n_1,n_2,n_3>0}{\sss n_1+n_2-n_3=0}}
(\al_{-n_3}\al_{+n_1}\al_{+n_2}+\al_{-n_1}\al_{-n_2}\al_{+n_3})\\
A_{3n^2}&=&\sum_{\tl{\sss n_1,n_2,n_3>0}{\sss n_1+n_2-n_3=0}}
\frac{n_1^2+n_2^2+n_3^2}{3}
(\al_{-n_3}\al_{+n_1}\al_{+n_2}+\al_{-n_1}\al_{-n_2}\al_{+n_3})\nonumber\\ 
A_4&=&\sum_{\tl{\sss n_1,n_2,n_3,n_4>0}{\sss \sum \pm n_i=0}}
\Big(\al_{-n_1}\al_{-n_2}\al_{-n_3}\al_{+n_4}+\nonumber\\ 
&+&\frac{3}{2}\al_{-n_1}\al_{-n_2}\al_{+n_3}\al_{+n_4}
+\al_{-n_1}\al_{+n_2}\al_{+n_3}\al_{+n_4}\Big) \nonumber\\ 
A_5&=&\sum_{\tl{\sss n_1,\dots,n_5>0}{\sss \sum \pm n_i=0}}
(\al_{-n_1}\al_{-n_2}\al_{-n_3}\al_{-n_4}\al_{+n_5}+ \nonumber\\
&+&2\al_{-n_1}\al_{-n_2}\al_{-n_3}\al_{+n_4}\al_{+n_5}
+2\al_{-n_1}\al_{-n_2}\al_{+n_3}\al_{+n_4}\al_{+n_5}+\nonumber\\
&+&\al_{-n_1}\al_{+n_2}\al_{+n_3}\al_{+n_4}\al_{+n_5})\nonumber\,, 
\eeqa
where by $\sum \pm n_i$ we mean, for each term, 
the sum of the indexes with their signs, for example for the first
term in $A_4$, $\sum \pm n_i=-n_1-n_2-n_3+n_4$.

The last ingredients are the substitution for the Virasoro zero modes 
in terms of $\al_0$, ${\cal N}$ and $\la$
\beq
\label{virasoro-zero}
L_0  =  \frac{1}{2} \alpha_0^2 + \mathcal{N} - \(\lambda -
\frac{1}{2}\) \alpha_0\,,
\eeq
and the substitution for the constraint \refeq{vinc-bos} that,
introducing $w$ as in \cite{ym2}, we can write as
\beq
\label{alpha0sub}
\al_0=\la-1+\frac{w}{2},\qquad\qquad
\Tilde \al_0=\la-1-\frac{w}{2}\,.
\eeq
The result for the $SU(N)$ hamiltonian must be independent of $w$ (see
\cite{ym2}): we check that this is indeed the case.

The final form for the string hamiltonian in the generalized YM with
quartic Casimir is 
\beqa
H&=&\la_4 L\Bigg[ \frac{{\cal N}+\Tilde {\cal N}}{6}+\frac{7}{6 N}(A_3+\Tilde
A_3)+\nonumber\\
&+& \frac{1}{N^2}\bigg(A_{2n^2}+ \Tilde A_{2n^2} +2(A_4+\Tilde
A_4)-\frac{13}{6}({\cal N}-\Tilde {\cal N})^2\bigg)+\nonumber \\
&+&\frac{1}{N^3}\bigg(\frac{3}{2}(A_{3n^2}+ \Tilde A_{3n^2})-6({\cal
N}-\Tilde {\cal N})(A_3-\Tilde A_3)+A_5+\Tilde A_5\bigg)+\nonumber\\
&+&\frac{6({\cal
N}-\Tilde {\cal N})^2 (A_3+\Tilde A_3)}{N^5}-\frac{3({\cal N}-\Tilde {\cal N})^4}{N^6}\Bigg]\,.
\eeqa

As we can see, the final result is independent of $w$ 
and $\la$. This is a non-trivial test
of the correctness of the result, as the two parameters were present
in the intermediate expressions in a very complex way.

To interpret this result first of all we recall that, as in the
standard case, the powers of $g_s=\frac{1}{N}$ correspond to the
factors $g_s^{-\chi}$ where $\chi$ is the Euler characteristic of the
surface involved. 

As in the standard quadratic Casimir case, we see that this
hamiltonian presents a simple free propagation part, ${\cal N}+\Tilde
{\cal N}$ and an interaction  vertex $V_3=\frac{1}{N}(A_3+\Tilde A_3)$ 
that connects three  
strings (two string joining into one, or one splitting in two),
conserving the total winding number and the orientation. 

In this generalized theory there also new vertices
$V_4=\frac{1}{N^2}(A_4+\Tilde A_4)$ and 
$V_5=\frac{1}{N^3}(A_5+\Tilde A_5)$ that simply connect four and five
strings, still 
conserving the total winding number and the orientation. The power of
$N$ is the correct $g_s^{-\chi}$ factor.

The term $\frac{1}{N^2}({\cal N}-\Tilde {\cal N})^2$ is the same one has in the
standard theory, and it is interpreted as usual as a contribution from
microscopic tubes and handles connecting the sheets of each string or
the sheets of two different strings. The contribution to the Euler
characteristic for the addition of a tube or of
an handle is $-2$, so the $1/N^2$ factor is
correct. In the same way $\frac{1}{N^6}({\cal N}-\Tilde {\cal N})^4$
can be interpreted as the contribution from a
microscopic surface, topologically identified with a sphere with 4
holes, that connects 4 sheets. The $-6$ change of the
Euler characteristic given by the connection of the sheets with this surface 
gives the factor $1/N^6$.

We have also other terms with a less clear geometric
interpretation; 
for example we can interpret the terms $\frac{1}{N^3} ({\cal N}-\Tilde
{\cal N}) (A_3-\Tilde A_3)$ and $\frac{1}{N^5} ({\cal N}-\Tilde {\cal N})^2
(A_3+\Tilde A_3)$ as combinations of a cubic vertex and handles.
Although this interpretation is intricated and even not unique, this
is not a limitation, since the important thing is that we have found
an hamiltonian which resumes in a compact way all the perturbative
aspects of the string theory.

\section{Conclusions}

In this paper we have investigated the two-dimensional generalized Yang
Mills theory from the point of view of the corresponding string
description and in particular the calculation of the string
hamiltonian. 

We have focused only on the quartic Casimir case, but from this example we can
extract some general features of the string description for the
generic theory.

The string interpretation is similar to the standard case: the Hilbert
space of the theory is the same (states of strings winding around the
spatial direction) and the hamiltonian has a similar structure, it
simply contains new kind of interaction vertices.

In particular for a Casimir of order $k$ one obtains all the vertices
connecting up to $(k + 1)$ strings; moreover in general there are many
other interaction terms which are combinations of vertices and
microscopic tubes
or handles, for which we haven't found a general rule: it is not clear
why for a specified Casimir there are particular terms and not
others. But in general we can say that for a generic combination of
higher Casimirs we expect to obtain a generic combination of the
possible interaction terms.

Thus the string field theory hamiltonian summarizes quite compactly 
the perturbative features of the string description.

\bigskip \bigskip
\noindent
{\bf Acknowledgments}

\medskip

We would like to thank Prof. Michele Maggiore for having followed our
work very closely, and for his suggestions.

\vspace{1cm}

\appendix

\section{The expression of the $\gamma_k$'s in terms of the $l_i$'s}
\label{gamma in l}

In this appendix 
we present the calculation that we have done to write down the values
of the gamma terms, and in particular the list \refeq{gammalist}.
The gamma terms are
\beq
\gamma_k\equiv \frac{\chi_R(P_k)}{d_R}=k\frac{\chi_R(T_k)}{d_R}\,.
\eeq
Following Hamermesh \cite{hamer} (Sect.~7-6, where $h_i$ is our $l_i$)
we obtain a formula for $\chi_R(T_k)/d_R$:
\beq
k\frac{\chi_R(T_k)}{d_R}=\frac{\sum_i l_i(l_i-1)\cdots
(l_i-(k-1))D[l_1,\dots,l_{i-1},l_i-k,l_{i+1},\dots l_N]}{D[l_1,l_2\dots l_N]}\,,
\eeq
where $D[l_1,l_2\dots l_N]\equiv \prod_{i<j} (l_i-l_j)$ 

We reduce this expression to a polynomium of $l_i$ using the following formula:
\beqa
\label{formula1}
F_x[l_i^z]\equiv\frac{\sum_i l_i^z D[l_1,\dots,l_{i-1},l_i+x,l_{i+1},\dots
l_N]}{D[l_1,l_2\dots l_N]}=\sum_i l_i^z \prod_{j\neq
i}\(1+\frac{x}{l_i-l_j}\)=\nonumber\\ 
=\sum_{a=0}^{z} x^a \!\!\!\sum_{\begin{array}{c}\sss
b_1,b_2..b_{a+1}\in\{1..N\}\\ \sss b_1<..<b_{a+1}\end{array}}
E_{z,a+1}[l_{b_1},\dots, l_{b_{a+1}}]\,,
\eeqa
where
\beq
\label{formula2}
E_{z,m}[x_1,x_2,\dots,x_m]=\sum_i\frac{x_i^{\,z}}{\prod_{j\neq i}
(x_i-x_j)}=\sum_{\begin{array}{c} \sss a_1,a_2\dots a_m\geq 0\\ \sss \sum_j
a_j=z-m+1\end{array}} x_1^{a_1} x_2^{a_2} \cdots x_m^{a_m}\,.  
\eeq
We postpone the demonstration of these formulas at the end of the appendix. 
We use the formulas above to calculate
 $\gamma_1$, $\gamma_2$, $\gamma_3$ and $\gamma_4$, that have
the following simple expressions in terms of $F_x[l_i^\beta]$:
\begin{itemize}
\item 
$\gamma_1$ is $F_x[l_i]$ for $x=-1$
\item
$\gamma_2$ is $F_x[l_i(l_i-1)]=F_x[l_i^2-l_i]=F_x[l_i^2]- F_x[l_i]$ for $x=-2$.
\item
$\gamma_3$ is $F_x[l_i(l_i-1)(l_i-2)]=F_x[l_i^3-3
l_i^2+2 l_i]=F_x[l_i^3]-3 F_x[l_i^2]+2 F_x[l_i]$ for $x=-3$.
\item
$\gamma_4$ is $F_x[l_i(l_i-1)(l_i-2)(l_i-3)]=F_x[l_i^4-6
l_i^3+11 l_i^2-6 l_i]=F_x[l_i^4]-6 F_x[l_i^3]+11 F_x[l_i^2]+-6
F_x[l_i]$ for $x=-4$.
\end{itemize}

We have (notice that $E_{a,a+1}[\dots]=1$)
\beq
F_x[l_i]=\sum_i E_{1,1}[l_i]+x\sum_{i<j} E_{1,2}[l_i,l_j]=\sum_i l_i+x
\binom{N}{2}\,, 
\eeq
\beqa
F_x[l_i^2]&=&\sum_i E_{2,1}[l_i]+x\sum_{i<j} E_{2,2}[l_i,l_j]+x^2
\sum_{i<j<k} E_{2,3}[l_i,l_j,l_k]=\nonumber\\
&=&\sum_i l_i^2+x (N-1)\sum_i l_i+x^2\binom{N}{3}\,,
\eeqa
\beqa
F_x[l_i^3]&=&\sum_i E_{3,1}[l_i]+x\sum_{i<j} E_{3,2}[l_i,l_j]+x^2
\sum_{i<j<k} E_{3,3}[l_i,l_j,l_k]+\nonumber \\
&+&x^3\sum_{i<j<k<m} E_{3,4}[l_i,l_j,l_k,l_m]=\nonumber\\
&=&\sum_i l_i^3+x((N-1)\sum_{i} l_i^2+\sum_{i<j} l_i l_j)+x^2
\binom{N-1}{2}\sum_{i} l_i+x^3\binom{N}{4}\,,
\eeqa
\beqa
F_x[l_i^4]&=&\sum_i E_{4,1}[l_i]+x\sum_{i<j} E_{4,2}[l_i,l_j]+x^2
\sum_{i<j<k} E_{4,3}[l_i,l_j,l_k]+\nonumber \\
&+&\!\!x^3\sum_{i<j<k<m}
E_{4,4}[l_i,l_j,l_k,l_m]+x^4\binom{N}{5}=\nonumber\\ 
&=&\sum_i l_i^4+x\Big[(N-1)\sum_i l_i^3+x\sum_{i<j}(l_i^2 l_j+l_i
l_j^2)\Big]+x^2\Big[\Big(\!\!\tl{N-1}{2}\!\!\Big)\sum_i
l_i^2+\nonumber\\
&+&\!\!(N-2)\sum_{i<j} l_i l_j\Big]+x^3\binom{N-1}{3}\sum_i
l_i+x^4\binom{N}{5}\,. 
\eeqa

After some calculation, we obtain the list
\beqa
\gamma_1&=&\sum_i l_i-\binom{N}{2}=\sum_i n_i\quad=C_1=n\nonumber\\
\gamma_2&=&\sum_i l_i^2-(2N-1)\sum_i l_i+\frac{N(N-1)(2N-1)}{3}\nonumber\\
\gamma_3&=&\sum_i l_i^3-3 N \sum_i l_i^2-3 \sum_{i<j}l_i l_j+
\frac{9N^2-9N+4}{2} \sum_i l_i+\nonumber\\
&-&\frac{3}{8}N(N-1)(3N^2-3N+2)\\
\gamma_4&=&\sum_i l_i^4-4 \sum_{i<j} (l_i^2 l_j+l_i
l_j^2)-(4N+2)\sum_i l_i^3+ (16N-8) \sum_{i<j}l_i l_j+ \nonumber\\
&+&(8N^2+3)\sum_i
l_i^2-\frac{2}{3}(2N-1)(8N^2-8N+9) \sum_i l_i+\nonumber\\
&+&\frac{4}{7}N(N-1)(2N-1)(4N^2-4N+7)\nonumber\,,
\eeqa 
that we use in Sect.~\ref{calc-higher}.

\vspace{0.8cm}

In the following we demonstrate the formulas \refeq{formula1},
\refeq{formula2}. We recall the definitions and the formulas to be proven:
 
{\bf Definition 1:}
\beq
F_x[l_i^z]\equiv\frac{\sum_i l_i^z D[l_1,\dots,l_{i-1},l_i+x,l_{i+1},\dots
l_N]}{D[l_1,l_2\dots l_N]}\,.
\eeq

{\bf Definition 2:}
\beq
E_{z,m}[x_1,x_2,\dots,x_m]\equiv \sum_i\frac{x_i^{\,z}}{\prod_{j\neq i}
(x_i-x_j)}\,.
\eeq

{\bf Theorem 1:}
\beqa
F_x[l_i^z]=\sum_i l_i^z \prod_{j\neq
i}\(1+\frac{x}{l_i-l_j}\)=\nonumber\\ 
=\sum_{a=0}^{z} x^a \!\!\!\sum_{\begin{array}{c}\sss
b_1,b_2..b_{a+1}\in\{1..N\}\\ \sss b_1<..<b_{a+1}\end{array}}
E_{z,a+1}[l_{b_1},\dots, l_{b_{a+1}}]\,.
\eeqa

{\bf Theorem 2:} 
\beq
E_{z,m}[x_1,x_2,\dots,x_m]=\sum_{\begin{array}{c} \sss a_1,a_2\dots
a_m\geq 0\\ \sss \sum_j a_j=z-m+1\end{array}} x_1^{a_1} x_2^{a_2}
\cdots x_m^{a_m}\,.   
\eeq

The two theorems give the desired formulas. 

{\bf Proof of Theorem 1:}

From the Definition 1 of $F_x[..]$:
\beqa
F_x[l_i^z]
&=&\sum_i l_i^z \prod_{a,\; (a<i)} \(1-\frac{x}{l_a-l_i}\)
\prod_{b,\; (b>i)} \(1+\frac{x}{l_i-l_b}\)=\nonumber\\
&=& \sum_i l_i^z \prod_{j,\; j\neq i} \(1+\frac{x}{l_i-l_j}\)=\nonumber\\   
&=& l_1^z\(1+\frac{x}{l_1-l_2}\)\(1+\frac{x}{l_1-l_3}\)\cdots 
         \(1+\frac{x}{l_1-l_N}\)+ \\
&+& l_2^z\(1+\frac{x}{l_2-l_1}\)\(1+\frac{x}{l_2-l_3}\)\cdots 
         \(1+\frac{x}{l_2-l_N}\)+\dots\nonumber \\
\dots &+& l_N^z\(1+\frac{x}{l_N-l_1}\)\(1+\frac{x}{l_N-l_3}\)\cdots 
         \(1+\frac{x}{l_N-l_{N-1}}\)=\nonumber\,
\eeqa
then we can collect the powers of $x$ easily:
\beqa
&=&\sum_b l_b^z+ x\sum_{b_1<b_2}\sum_{i=1}^2
\frac{l_{b_i}^z}{\prod_{j=1..2,\;j\neq
i}(l_{b_i}-l_{b_j})}+x^2\sum_{b_1<b_2<b_3}\sum_{i=1}^3 
\frac{l_{b_i}^z}{\prod_{j=1..3,\;j\neq i}(l_{b_i}-l_{b_j})}+\dots\nonumber\\
&+&x^a\sum_{b_1<b_2<b_3<..<b_{a+1}}\sum_{i=1}^{a+1}
\frac{l_{b_i}^z}{\prod_{j=1..(a+1),\;j\neq
i}(l_{b_i}-l_{b_j})}+\dots\\
&+&x^{N-1}\sum_{b_1<b_2<b_3<..<b_{N}}\sum_{i=1}^{N}
\frac{l_{b_i}^z}{\prod_{j=1..N,\;j\neq i}(l_{b_i}-l_{b_j})}=\nonumber\,
\eeqa
in terms of $E_{z,m}$ previously defined:
\beq
=\sum_a^{N-1} x^a \sum_{b_1<b_2<..<b_{a+1}=1..N} E_{z,a+1}[l_{b_1},l_{b_2},\dots,l_{b_{a+1}}]\,,
\eeq
and this is Theorem 1.

{\bf Proof of Theorem 2:}

This identity is very simple for $m=2$:
\beq
E_{z,2}=\frac{x_1^z}{x_1-x_2}+\frac{x_2^z}{x_2-x_1}=
\frac{x_1^z-x_2^z}{x_1-x_2}=x_1^{z-1}+x_2^{z-1}+x_1^{z-2}x_2+x_1^{z-3}x_2^2+\dots+x_1 x_2^{z-2}\,. 
\eeq

The proof is by induction. Defining
\beq
E'_{z,m}=\sum_{\begin{array}{c} \sss a_1,a_2\dots
a_m\geq 0\\ \sss \sum_j a_j=z-m+1\end{array}} x_1^{a_1} x_2^{a_2}
\cdots x_m^{a_m}\,,   
\eeq
the hypothesis of the induction is 
\beq
E_{z,i}=E'_{z,i} \quad \forall i\in\{2,3,\dots,m-1\}\quad \textrm{and}
\quad \forall z \in I\!\!N\,,
\eeq
and the thesis of the theorem is 
\beq
E_{z,m}=E'_{z,m} \quad \forall z \in I\!\!N\,.
\eeq
Let us start from $E'_{z,m}$ and try to write it in terms of $E_{z,m}$
using the hypothesis of the induction for $m-1$, isolating one of the
$x$ (it does not matter which $x$, we isolate $x_m$ and use the
hypothesis on $x_1,x_2,\dots x_{m-1}$, but they are all equivalent):  
\beq
E'_{z,m}=\sum_{\begin{array}{c} \sss a_1,a_2\dots a_{m-1},
a_m\geq 0\\ \sss \sum_{j=1}^{m-1} a_j=(z-1)-a_m-(m-1)+1\end{array}}
x_1^{a_1} x_2^{a_2}\cdots x_m^{a_m}\,.
\eeq
Using the hyp.:
\beqa
E'_{z,m}&=&\sum_{a_m\geq 0}
\sum_{i=1}^{m-1}\frac{x_i^{z-1-a_m}}{\prod_{j=1,\dots,m-1;\; j\neq
i}(x_i-x_j)} x_m^{a_m}=\nonumber\\ 
&=& \sum_{i=1}^{m-1}\[\frac{x_i^z}{\prod_{j=1,\dots,m;\; j\neq i}(x_i-x_j)}+
\frac{x_m^z}{\prod_{j=1,\dots,m-1;\; j\neq i}(x_i-x_j)(x_m-x_j)}\]\,,
\eeqa
where we have used the formula for $m=2$ that is simply:
\beq
\sum_{a_m\geq 0} x_i^{z-1-a_m} x_m^{a_m}=\frac{x_i^z}{x_i-x_m}+\frac{x_m^z}{x_m-x_i}\,.
\eeq 
Now we have:
\beqa
E'_{z,m}&=&\sum_{i=1}^{m}\frac{x_i^z}{\prod_{j\neq i}^{m}(x_i-x_j)}+\nonumber\\
&+&x_m^z
\[\sum_{i=1}^{m-1}\frac{1}{\prod_{j\neq i}^{m-1}(x_i-x_j)(x_m-x_i)}-   
\frac{1}{\prod_{j\neq m}^{m}(x_m-x_j)}\]=\nonumber\\
&=&E_{z,m}-x_m^z \[\sum_{i=1}^{m-1}\frac{1}{\prod_{j\neq i}^{m}(x_i-x_j)}+   
\frac{1}{\prod_{j\neq m}^{m}(x_m-x_j)}\]=\nonumber\\
&=&E_{z,m}-x_m^z \sum_{i=1}^{m}\frac{1}{\prod_{j\neq
i}^{m}(x_i-x_j)}=E_{z,m}-x_m^z E_{0,m}\,.
\eeqa
Thus we have
\beq
E'_{z,m}=E_{z,m}-x_m^z E_{0,m}\,.
\eeq
Notice that we know that $E_{0,i}=0$ only for $i$ up to $m-1$ from the
induction hypothesis. But now we use the fact that we can isolate a
generic $x$: the actual formula is
\beq
E'_{z,m}=E_{z,m}-x_j^z E_{0,m}\quad \forall j\in \{1,2\dots,m\}\,,
\eeq
thus if we consider this for $j=a$ and $j=b$ with $a\neq b$ and take
the difference we have:   
\beq
(x_a^z-x_b^z)E_{0,m}=0 \quad \textrm{valid for generic $x$} \quad
\Rightarrow E_{0,m}=0\,.
\eeq
Finally we have the thesis
\beq
E'_{z,m}=E_{z,m}\,.
\eeq

\end{document}